\documentclass{mem}
\usepackage{natbib}\usepackage{txfonts}\usepackage{balance}
\usepackage{graphicx}
\usepackage[a4paper,breaklinks,dvipdfm]{hyperref}
\idline{00}{000}
\begin{document}
\def \teff{$T\rm_{eff}$}
\def \kms{$\mathrm {km s}^{-1}$}
\newcommand{\ssun}{\ensuremath{_\odot}}
\newcommand{\feoh}{\ensuremath{[\mathrm{Fe/H}]}}

\title{Light element abundances in the Galactic globular cluster 47 Tuc}

\author{ A.\,\v{C}erniauskas\inst{1}, A.\,Ku\v{c}inskas\inst{1}, P.\,Bonifacio\inst{2}, S. M.\, Andrievsky\inst{3}, S. A.\,Korotin
\inst{3},
		 V.\,Dobrovolskas\inst{4}}
		
\offprints{A. \v{C}erniauskas}

\institute{\inst{1} Institute of Theoretical Physics and Astronomy, Vilnius
           University, A. Go\v stauto 12, Vilnius, LT--01108, Lithuania,
					\email{algimantas.cerniauskas@tfai.vu.lt} \\
           \inst{2} GEPI, Observatoire de Paris, CNRS, Universit\'{e} Paris Diderot, Place Jules Janssen, 92190, Meudon, France \\
           \inst{3}Department of Astronomy and Astronomical Observatory, Odessa National University and Isaac Newton Institute of Chile Odessa 						 branch,  Shevchenko Park, 65014 Odessa, Ukraine \\
           \inst{4} Astronomical Observatory, Vilnius University, M.~K.~\v{C}iurlionio 29, Vilnius, LT--03100, Lithuania \\
          }
\authorrunning{A.\,\v{C}erniauskas}
\titlerunning{Light element abundances in 47 Tuc}

\abstract{It is very likely that most (perhaps all) Galactic globular clusters (GGCs) have experienced two or even more star-formation episodes. This is indicated, in particular, by peculiar chemical composition of the cluster stars which show large variation in the abundances of light elements, such as Li, C, N, O, Na, Mg, and Al. We studied the abundances of Na, Mg, and Al in the atmospheres of 103 red giant branch stars located below (49 stars) and above (54 stars) the RGB bump in the GGC 47 Tuc. Our results show that the spread of [Na/Fe] abundance ratios is about three times larger than that of [Mg/Fe]. Our data also confirm the existence of weak Na-Al correlation, similar to the one observed in other GGCs. At the same time, we find no evidence for the existence of three populations of stars characterized with different abundances of aluminum, as reported recently by \citet{Ca13}.

}

\maketitle{}

\section{Introduction}

Galactic globular clusters (GGCs) are old, metal-poor stellar systems which for many years were thought to be good examples of simple stellar populations. However, recent photometric and spectroscopic observations seem to support the idea that stars in these stellar systems could have formed during two or more star formation episodes. This scenario is corroborated by the fact that abundances of light elements (Li, O, Na, Mg, Al) in the GGCs show various correlations/anti--correlations, such as Na--O anti--correlation \citep{Kr94}, Li--O correlation \citep{Pa05}, Li--Na anti--correlation \citep{Bo07}, Mg--Al anti--correlation \citep{Ca09}. It is important to mention that such abundance trends are not seen in the halo field stars of the same metallicity.

It is generally accepted today that stars characterized by different chemical composition belong to at least two different stellar generations: second generation stars (e.g., O-poor, Na-rich) are considered to be younger than first generation stars (O-rich, Na-poor), and therefore, must have formed from the matter enriched in chemical elements which were synthesized by the first generation stars \citep[see, e.g.,][]{Gr01}. It is thought that two types of first generation stars could have been responsible for the enrichment of second generation stars: (i) AGB stars \citep{Ve01}, or (ii) fast rotating massive stars \citep{De07}. However, none of the two scenarios can fully account for the abundance trends observed in the GGCs. Therefore, further insights about the chemical evolution of the GGCs based on the results of homogeneous chemical analysis of large number of stars in multiple GGCs, as well as new theoretical interpretations, would be highly desirable. Galactic globular cluster 47 Tuc is particularly interesting target in this respect, due to its proximity and large angular diameter.

In this contribution we present first results of our study aimed to investigate the
abundance trends of Na, Mg, and Al in 47 Tuc, to better constrain the possible chemical evolutionary scenarios of this GGC.

\begin{figure}
\begin{center}
\includegraphics[scale=0.60]{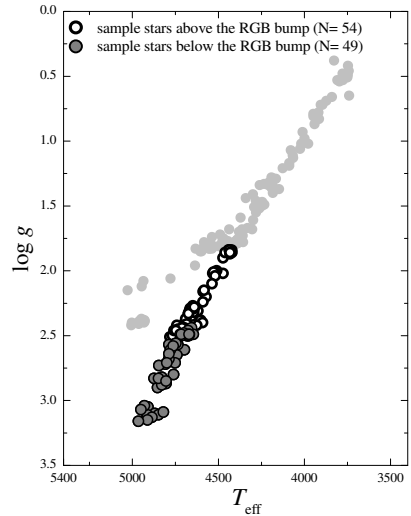}
\caption{\footnotesize $T\rm_{eff}$--log \textit{g} diagram of 47 Tuc showing RGB stars studied in this work.}
\label{HR}
\end{center}
\vspace{-5mm}
\end{figure}

\section{Observations and data reduction}

We studied 103 red giant branch (RGB) stars below (49 stars) and above (54 stars) the RGB bump in 47 Tuc using archival spectra retrieved from the ESO Science Archive Facility (program 072.D-0777). Spectra were acquired with the multifibre FLAMES/GIRAFFE spectrograph mounted on the VLT UT2 telescope. Observations were done using three GIRAFFE setups: HR~13 ($\Delta\lambda=612-640$~nm, typical $S/N\approx75$), HR~15 ($\Delta\lambda=660-696$~nm, $S/N\approx110$), and HR~21  ($\Delta\lambda=848-900$~nm, $S/N\approx160$). Spectral resolution at the central wavelength of these three setups was \textit{R} = 22500, 19300, and 16200, respectively. Exposure times were 1600s and 3600 s for stars in the magnitude ranges of 11$<$\textit{V}$<$13 and 13$<$\textit{V}$<$16, respectively.

The spectra were bias-corrected, flat-fielded and then wavelength-calibrated using GIRAFFE pipeline package {\tt ESORex} \footnote{https://www.eso.org/sci/facilities/paranal\\ /instruments/flames/doc/VLT-MAN-ESO-13700-4034v85.pdf }. During the original observation run no fibers were dedicated to observe the sky. Therefore, to estimate the contribution of the sky we used archival sky observations made with the same GIRAFFE gratings and carried out at the zenith angle, Moon phase, and with exposure times very similar to those in the original observations of our sample RGB stars (programs 079.B-0721, 071.D-0065, and 082.B-0940 for HR~13, HR~15, and HR~21, respectively). We found that the sky  intensity was negligible and it has never exceeded 1\% of the continuum intensity in the spectra of faintest stars in our sample, therefore, we did not subtract sky from the spectra of our sample stars. Continuum normalization procedure was done using the IRAF \citep{To86} \textit{continuum} task. Radial velocities were measured using the IRAF \textit{fxcorr} task, to make sure that all stars are cluster members.

\section{Abundance determination}

Effective temperatures of the sample stars were determined using photometric \textit{V--I} observations of 47 Tuc made by \citet{Ber09} and $T\rm_{eff}$--\textit{(V--I)} calibrations of \citet{RaMe93}, while their surface gravities were estimated using the classical relation between the surface gravity, mass, effective temperature, and luminosity (Fig.~\ref{HR}).

Abundances of Na, Mg, and Al were derived under the assumption of LTE, by utilizing classical one-dimensional (1D) hydrostatic ATLAS9 model atmospheres \citep{Kur93,Sbr05}. Model atmospheres were computed using the mixing length parameter $\alpha_{\rm MLT}=1.25$, with the overshooting switched off. SynthV spectral synthesis code \citep{Ts96} was used to compute spectral line profiles of Na, Mg, and Al in LTE. We adopted a constant metallicity value of $\feoh=-0.7$ \citep{Mc11} for all sample stars, while the microturbulence velocities, $\xi_{\rm t}$, were estimated using empirical calibration of \citet{Jo08}. Atomic parameters of spectral lines used in the abundance derivations were taken from the VALD database \citep{Ku99} and are provided in Table~\ref{param}.

Abundance sensitivity to uncertainties in the atmospheric parameter determination was estimated using three sets of values of $T\rm_{eff}$ and $\log g$, two of which were bracketing the extreme values of $T\rm_{eff}$ and $\log g$ in our sample stars, and one was corresponding to a typical RGB bump star. These atmospheric parameters were then varied by the following values: $\Delta T\rm_{eff}$=100 K, $\Delta\log g$ = 0.2 and $\Delta\xi_{\rm t}$=0.2\,km~s$^{-1}$. The resulting changes in the obtained elemental abundances are listed in Table~\ref{err}.

\begin{table}
\caption{Atomic parameters of spectral lines used in the abundance determinations of Na, Mg, and All. Natural, Stark, and van der Waals broadening constants are provided in the last three columns.}
\label{param}
\vspace{-5mm}
\begin{center}
\scalebox{0.95}{
\setlength{\tabcolsep}{2pt}
\begin{tabular}{lcccccc}
\hline
\hline
Element & $\lambda$, nm & $\chi$, eV & log$\textit{gf}$ & log $\gamma_{rad}$ & log$\frac{\gamma_4}{N_e}$ & log$\frac{\gamma_6}{N_H}$ \\
\hline
Na I & 615.4 & 2.10 & $-1.56 $ & 7.85 & $-4.38$ & $-7.28$ \\
Na I & 616.0 & 2.10 & $-1.26 $ & 7.85 & $-4.38$ & $-7.28$ \\
Mg I & 631.8 & 5.10 & $-1.72 $ & 7.75 & $-3.87$ & $-7.08$ \\
Mg I & 880.6 & 4.34 & $-0.13 $ & 8.69 & $-4.91$ & $-7.40$ \\
Al I & 669.6 & 3.14 & $-1.34 $ & 8.48 & $-4.34$ & $-7.26$ \\
Al I & 877.2 & 4.02 & $-0.31 $ & 7.92 & $-4.18$ & $-6.67$ \\
\hline
\end{tabular}
}
\end{center}
\end{table}

\begin{table}
\begin{center}
\caption{Sensitivity of the derived abundances to changes in the atmospheric parameters ($\Delta T\rm_{eff}$=100 K, $\Delta\log g$ = 0.2 and $\Delta\xi_{\rm t}$=0.2\,km~s$^{-1}$).}
\begin{tabular}{lcccc}
\hline\hline
Element & $\Delta$ $T\rm_{eff}$ & $\Delta\log g$ & $\Delta\xi_{\rm t}$  \\ [0.5ex]
\hline
Na I & -0.15 & 0.01 & 0.02 \\
Mg I & -0.10 & 0.03 & 0.03 \\
Al I & -0.12 & 0.01 & 0.04 \\
\hline
\end{tabular}
\label{err}
\end{center}
\vspace{-5mm}
\end{table}

\begin{figure*}[t!]
\resizebox{\hsize}{!}{\includegraphics[clip=true]{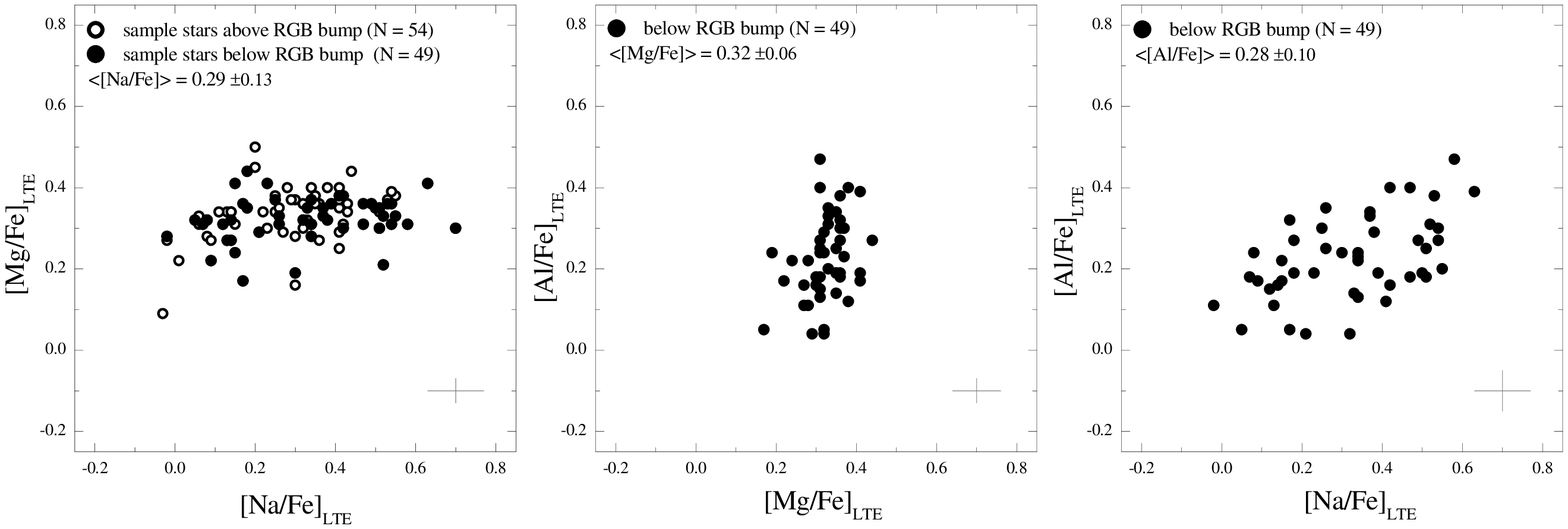}}
\caption{ [Mg/Fe] vs. [Na/Fe] (left panel), [Al/Fe] vs. [Mg/Fe] (middle panel), and
[Al/Fe] vs. [Na/Fe] (right panel) in the RGB stars of 47 Tuc, both below (filled circles) and above (open circles) the RGB bump. All abundance ratios are 1D LTE estimates. Typical error bars of the abundance determinations are shown in the lower right corner of each data panel. Note that in the middle and right panels the data are shown only for stars located below the RGB bump.}
\label{abn}
\end{figure*}

\section{Results and conclusions}

The following mean abundance--to--iron ratios were obtained for 103 RGB stars located both below and above the RGB bump: $\langle{\rm [Na/Fe]}\rangle=0.29\pm0.13$ and $\langle{\rm [Mg/Fe]}\rangle=0.32\pm0.12$ (square root of the abundance dispersion in all individual stars is provided after the $\pm$ sign). In case of Al, only stars below the RGB bump were studied so far (49 objects), resulting in $\langle$[Al/Fe]$\rangle$=0.28$\pm$0.13.

We find that [Mg/Fe] abundance ratios do not show large star--to--star variation in the RGB stars of 47 Tuc but the spread in [Na/Fe] ratios is more than two times larger than that in [Mg/Fe] (Fig.~\ref{abn}, \textit{left panel}). Moreover, we found no clear anti-correlation neither between  Na and Mg, nor between Mg and Al (Fig.~\ref{abn}). Nevertheless, it seems that our data suggests the existence of weak Na--Al correlation (Fig.~\ref{abn}, \textit{right}), similar to the one observed in
other GGCs \citep[see, e.g.,][]{Ca09}: the probability \textit{p} that this correlation does not exist (estimated using Kendall's $\tau$ test) is equal to $2.2 \times {10^{-4}}$, correlation coefficient is $0.51$.

It may be interesting to note that stars located above the RGB bump may display the signs of an additional 'thermohaline mixing' event \citep{Mu12}. However, we found no obvious differences in the abundance patterns of Na and Mg in our sample stars located below and above the RGB bump.

While the obtained results do not allow to unambiguously discriminate between the AGB and fast rotating massive star pollution scenarios, they do not seem to support the evidence of three distinct populations recently reported by \citet{Ca13}.

\begin{acknowledgements}
This work was supported by grants from the Research Council of Lithuania (MIP-065/2013, TAP LZ 06/2013). SMA and SAK acknowledge funding from the Research Council of Lithuania for the research visits to Vilnius. The study was based on observations made with the European Southern Observatory telescopes obtained from the ESO/ST-ECF Science Archive Facility.
\end{acknowledgements}

\bibliographystyle{aa}

\end{document}